\documentclass[%
reprint,
 amsmath,amssymb,
 aps,
]{revtex4-2}
\usepackage{graphicx}% Include figure files
\usepackage{dcolumn}% Align table columns on decimal point
\usepackage{bm}% bold math
\usepackage{float}
\usepackage{caption}
\usepackage{subcaption}
\usepackage{color}

\begin{document}

\preprint{APS/123-QED}

\title{Experimental evidence of nonlinear focusing in standing water waves}% Force line breaks with \\
%\thanks{A footnote to the article title}%

\author{Yuchen He$^{1}$} 
\email{yuchen.he@sydney.edu.au}
\author{Alexey Slunyaev$^{2,3}$}
\author{Nobuhito Mori$^{4}$} 
\author{Amin Chabchoub$^{5,4,1}$}
\email{chabchoub.amin.8w@kyoto-u.ac.jp}
\affiliation{$^1$ Centre for Wind, Waves and Water, School of Civil Engineering, The University of Sydney, Sydney, NSW 2006, Australia}
\affiliation{$^2$ National Research University-Higher School of Economics, 25 Bol’shaya Pechorskaya Street, Nizhny Novgorod 603950, Russia}
\affiliation{$^3$Institute of Applied Physics RAS, 46 Ulyanova Street, Nizhny Novgorod 603950, Russia}
\affiliation{$^4$ Disaster Prevention Research Institute, Kyoto University, Uji, Kyoto 611-0011, Japan} 
\affiliation{$^5$ Hakubi Center for Advanced Research, Kyoto University, Yoshida-Honmachi, Kyoto 606-8501, Japan}

%\collaboration{MUSO Collaboration}%\noaffiliation

%\author{Charlie Author}
% \homepage{http://www.Second.institution.edu/~Charlie.Author}
%\affiliation{
% Second institution and/or address\\
% This line break forced% with \\
%}%
%\affiliation{
% Third institution, the second for Charlie Author
%}%
%\author{Delta Author}
%\affiliation{%
% Authors' institution and/or address\\
% This line break forced with \textbackslash\textbackslash
%}%

%\collaboration{CLEO Collaboration}%\noaffiliation

\date{\today}% It is always \today, today,
             %  but any date may be explicitly specified

\begin{abstract}
Nonlinear wave focusing originating from the universal modulation instability (MI) is responsible for the formation of strong wave localizations on the water surface and in nonlinear wave guides, such as optical Kerr media and plasma. Such extreme wave dynamics can be described by breather solutions of the nonlinear Schr\"odinger equation (NLSE) like by way of example the famed doubly-localized Peregrine breathers (PB), which typify particular cases of MI. On the other hand, it has been suggested that the MI relevance weakens when the wave field becomes broadband or directional. Here, we provide experimental evidence of nonlinear and  distinct PB-type focusing in standing water waves describing the scenario of two counter-propagating wave trains. The collected collinear wave measurements are in excellent agreement with the hydrodynamic coupled NLSE (CNLSE) and suggest that MI can undisturbedly prevail during the interplay of several wave systems and emphasize the potential role of exact NLSE solutions in extreme wave formation beyond the formal narrowband and uni-directional limits. Our work may inspire further experimental investigations in various nonlinear wave guides governed by CNLSE frameworks as well as theoretical progress to predict strong wave coherence in directional fields.
%\begin{description}
%\item[Usage]
%Secondary publications and information retrieval purposes.
%\item[Structure]
%You may use the \texttt{description} environment to structure your abstract;
%use the optional argument of the \verb+\item+ command to give the category of each item. 
%\end{description}
\end{abstract}

%\keywords{Suggested keywords}%Use showkeys class option if keyword
                              %display desired
\maketitle

%\tableofcontents
The emergence of strongly localized waves in nonlinear dispersive media is an actively investigated field of research across wave physics \cite{onorato2013rogue,dudley2014instabilities,suret2020nonlinear,vanderhaegen2021extraordinary,weisman2021diffractive}. While excluding any external effects on a wave system, extreme wave formations in uni-directional wave guides can be explained either as a result of unstable nonlinear wave interaction \cite{zakharov1968stability,bonnefoy2016observation}, or linear superposition principle \cite{longuet1974breaking,mcallister2019laboratory}. Both mechanisms have been intensively studied in laboratory environments and the ocean \cite{dudley2019rogue,waseda2020nonlinear}. The modulation instability (MI) is a nonlinear wave focusing mechanism \cite{bespalov1966filamentary,benjamin1967disintegration}, which has been proven to be present in complex sea states, such as crossing seas \cite{waseda2009evolution, onorato2010freak,gramstad2018modulational}, however being less dominant compared to wave systems with a single wave vector due to the violation of both critical assumptions, namely uni-directionality and narrowband energy level \cite{janssen2003nonlinear,mori2011estimation}.

Then again, it has been recently conjectured that there is an increase of probability of rogue wave formation in coupled two-wave systems compared to an uncoupled directional wave field \cite{gronlund2009evolution}. A coupled nonlinear Schr\"odinger (CNLSE) framework has been already derived for water waves in the 80s \cite{okamura1984instabilities} and is nowadays considered as a fundamental dynamical system for the study of complex, coherent, directional, and rogue wave dynamics in various physical media \cite{onorato2006modulational,baronio2012solutions,frisquet2016optical,stole2018extreme,steer2019experimental}.

In this letter, we provide experimental evidence of nonlinear wave focusing in standing wave fields using the Peregrine breather (PB) \cite{peregrine1983water} as referenced nonlinear rogue wave and special MI model evolving in one of two counter-propagating wave systems. It is shown that Peregrine-type unsteady packets on finite and zero background can distinctly evolve in the presence of a counter-propagative regular wave field without any noticeable disturbance nor disintegration of coherence during wave focusing and defocusing. The experimental results are in excellent agreement with the hydrodynamic CNLSE. Moreover, we can confirm nonlinear focusing in such simplified but representative configuration by means of direct numerical simulations of the governing water wave equations using the high-order spectral method (HOSM) for potential flows.

The nonlinear interaction of standing waves involving an incident wave field characterized by the complex amplitude $\psi^{(1)}(x,t)$ with wavenumber $\kappa^{(1)}=(k,0)$ interacting with the reflected waves $\psi^{(2)}(x,t)$ with wavenumber $\kappa^{(2)}=(-k,0)$, and propagating along the space co-ordinate $x$, can be described by the CNLSE \cite{okamura1984instabilities,onorato2006modulational}
\begin{equation}\label{cnls}
    \begin{split}
    &i(\psi^{(1)}_x+\frac{1}{c_g}\psi^{(1)}_t)+\delta \psi^{(1)}_{tt}+\nu(|\psi^{(1)}|^2-2|\psi^{(2)}|^2)\psi^{(1)}=0.\\
    &i(\psi^{(2)}_x-\frac{1}{c_g}\psi^{(2)}_t)+\delta \psi^{(2)}_{tt}+\nu(|\psi^{(2)}|^2-2|\psi^{(1)}|^2)\psi^{(2)}=0,
    \end{split}
\end{equation}
The angular frequency $\omega^{(i)}$ and wavenumber $\kappa^{(i)}$ are connected through the deep-water linear dispersion relation $\omega^{(i)}=\sqrt{g\lvert\kappa^{(i)}\rvert}$, $i=1,2$ and $g$ being the gravitational acceleration, while the CNLSE parameters read
\begin{equation}\label{eqn_cnls_parameter}
    c_g=\frac{\omega}{2k},\ \delta=-\frac{k}{\omega^2},\ \nu=-k^3.
\end{equation} 
The standing water surface elevation describing the interaction between the incident wave field $\psi^{(1)}$ and opposing wave system $\psi^{(2)}$ to first-order of approximation is calculated by means of the complex envelope following the expression
\begin{equation}\label{eqn_elevation_unidirectional}
\begin{split}
    \eta(x,t)=\frac{1}{2}(&\psi^{(1)}(x,t)\exp[i(kx-\omega t)]\\ &+\psi^{(2)}(x,t)\exp[-i(kx+\omega t)]+c.c.).
\end{split}
\end{equation}
Here, $c.c.$ denotes the complex conjugation and we emphasize the weakening of the bound waves contributions in standing wave systems in contrast to the uni-directional wave evolution. Our wave modeling will be restricted to the CNLSE, described by the set of equations in (1), nonetheless, we would like to also highlight that a higher- and fourth-order CNLSE can be applied to two wave systems with different wave frequencies or directions \cite{gramstad2011fourth}.

We investigate the possibility of nonlinear wave focusing in a standing wave field, which is a special case of crossing wave systems with a crossing angle of $\pi$, by using the PB model as incident wave field $\psi^{(1)}$ and a regular wave envelope of same amplitude for $\psi^{(2)}$. We recall that the PB is a doubly-localized solution of the uni-directional nonlinear Schr\"odinger equation (NLSE) \cite{peregrine1983water}, which has been so far observed in optics, water waves and plasma \cite{kibler2010peregrine,chabchoub2011rogue,bailung2011observation}. That said, this solution describes the nonlinear stage of modulation instability in the case of infinitely long perturbation \cite{akhmediev1997solitons,osborne2010nonlinear} and has a rational growth. In fact, it can be approached by either Akhmediev- \cite{akhmediev1985generation,cavaleri2012rogue} or Kuznetsov breathers \cite{kuznetsov1977solitons}, and so far not discussed in the context of collinear waves. Its counterpart on zero background is also known as the degenerate soliton \cite{chabchoub2021peregrine}.

\begin{figure}[H]
\includegraphics[width=0.45\textwidth]{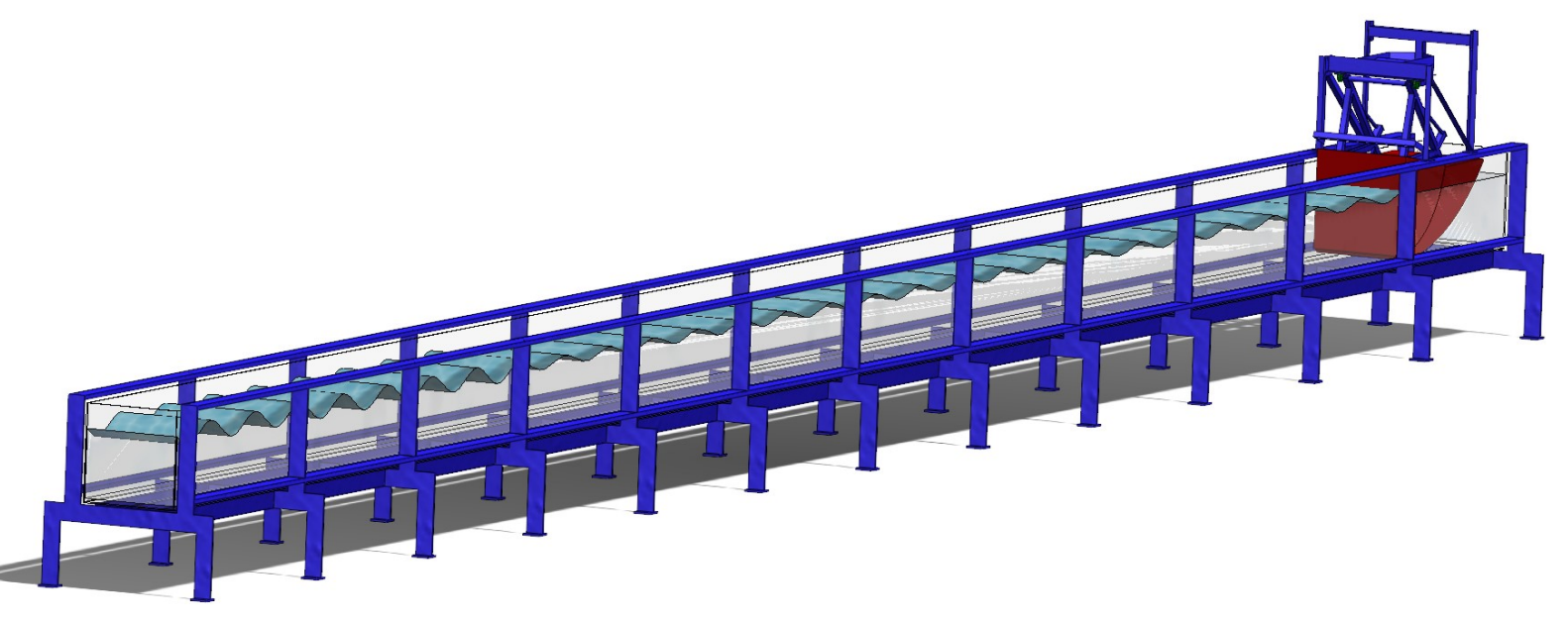}
\caption{Sketch of the wave flume setup as installed at the University of Sydney after removing the artificial beach installation to enable full wave reflection. The piston-type wave generator on the right end generates long-crested uni-directional waves. The glass wall and wave mirror is at the opposing end.}
\label{fig1}
\end{figure}

Since most of the state of the art wave flumes do not have two wave generators on opposing sides, one possibility to allow for wave train collisions is to make use of the wall at the opposite side of the wave generator as a reflective mirror for long-crested waves \cite{tanter1998influence}. Indeed, the absence of wave energy absorbing beach in a perfectly aligned flume permits a full reflection and thus, the formation of standing waves. Such adjusted and adopted set-up configuration can be viewed in Fig. \ref{fig1}.

The facility operates a piston-type wave paddle to generate boundary conditions in a form of surface elevation time-series. The tank has the dimensions $30\times 1\times 1$ m$^3$, the water depth is set to be $0.75$ m, and eight resistive wave gauges with a sampling rate of 32 Hz are used to measure the water surface elevation. The accurate controlled generation of the wave field allows the reiteration of the experiments with different wave gauge placements to ensure a high resolution also along the waves' propagation direction \cite{vanderhaegen2021extraordinary}.

\begin{figure}[H]
\begin{subfigure}{0.48\textwidth}
\includegraphics[width=1\textwidth]{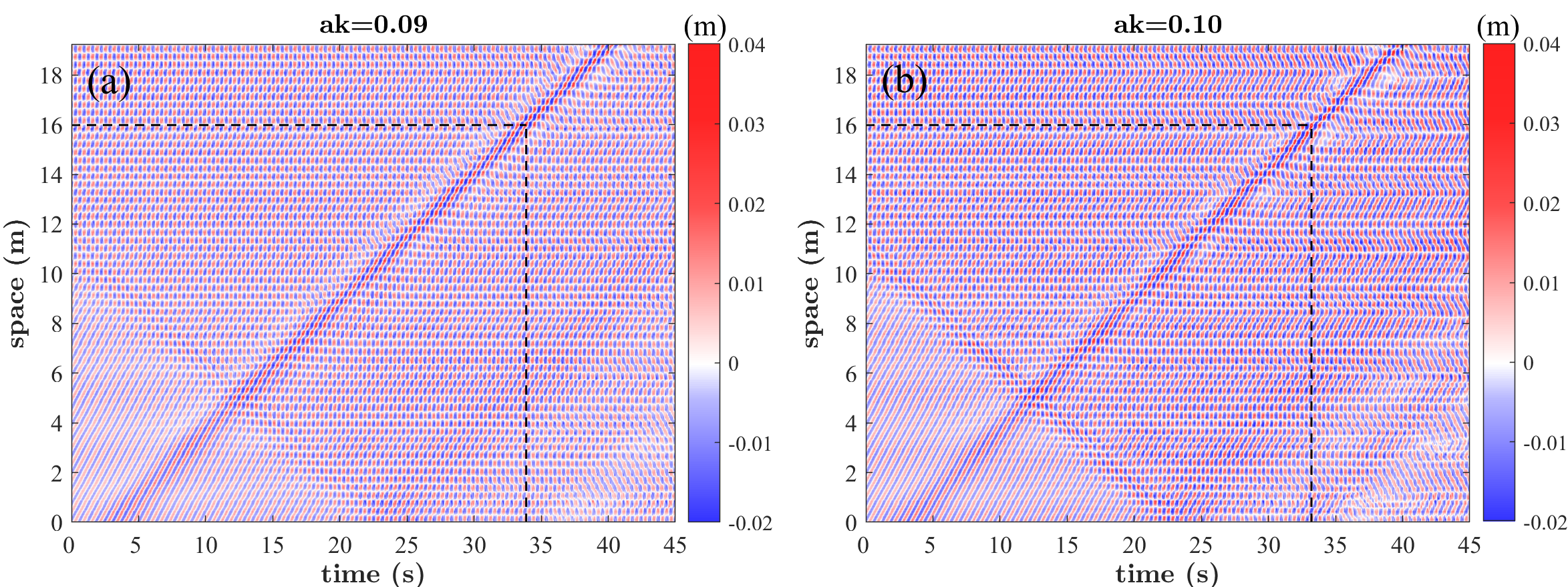}
\end{subfigure}
\begin{subfigure}{0.48\textwidth}
\includegraphics[width=1\textwidth]{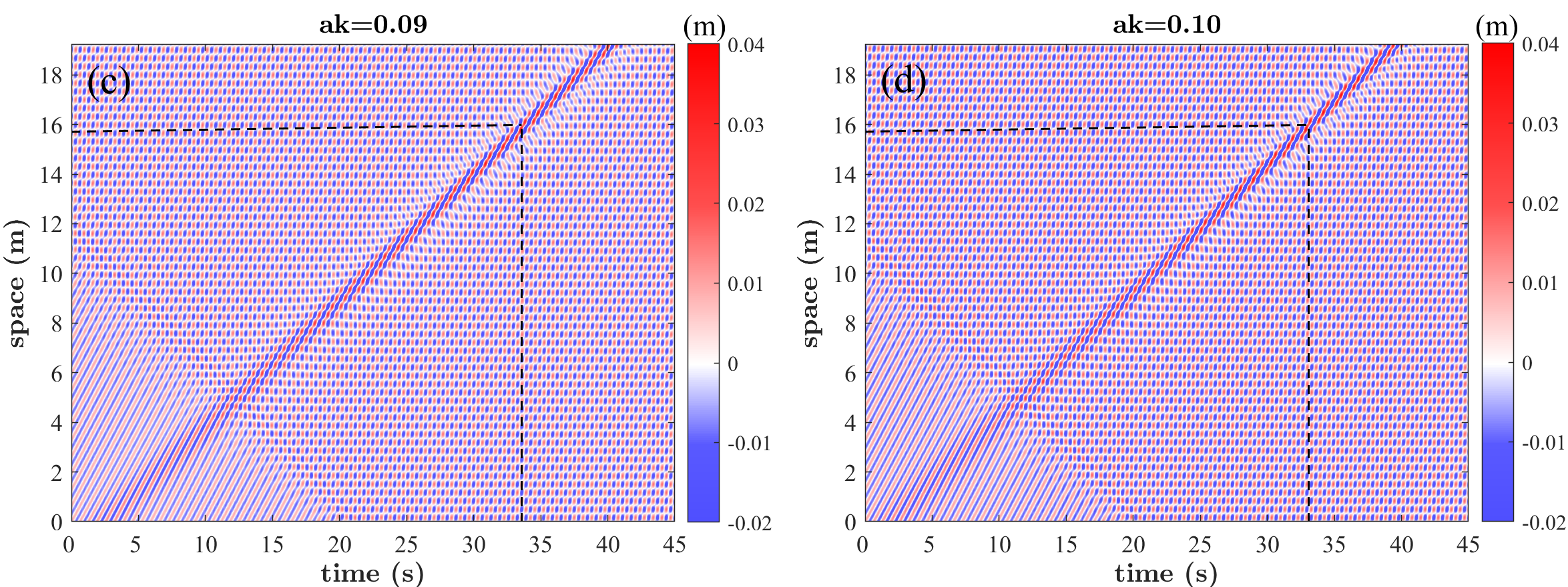}
\end{subfigure}
\caption{Top (a) and (b): Experimental observation of two standing PBs. Bottom (c) and (d): Numerical CNLSE simulations starting from the same boundary conditions as the laboratory experiments. Left (a) and (c): Wave parameters $a=0.01$ m and $ak=0.09$. Right (b) and (d): Wave parameters $a=0.01$ m and $ak=0.10$.}
\label{fig2}
\end{figure}

Since we are interested in the interaction of an incident PB wave field with a regular wave train with opposing wave vector, a dynamic head-on interaction which we will refer to as standing PB, we simply select a sufficiently long realization of the PB solution in time to guarantee the propagation of the unperturbed regular wave to the wall and back, before the wave maker launches the small and localized Peregrine-specific modulation. We refer to \cite{chabchoub2011rogue,chabchoub2016hydrodynamic} for the construction of the dimensional Peregrine boundary conditions. In our experiments we chose the amplitude to be $a=0.01$ m for two carrier steepness values $ak=0.09$ and $ak=0.10$. The respective values of the wave frequency can be computed using the linear dispersion relation $\omega=\sqrt{gk}$. The boundary conditions have been defined to expect the maximal wave compression 16 m from the wave generator. The results of the experimental campaign, accounting for 184 measurements along the longitudinal direction for each realization, together with the associated CNLSE predictions are shown in Fig. \ref{fig2}.

The numerical CNLSE simulations have been carried out for the same wave parameters and boundary conditions as the laboratory experiments for validations purposes. A pseudo-spectral approach and the fourth-order Runge–Kutta method have been adopted for the integration in space.  

\begin{figure}[H]
\includegraphics[width=0.23\textwidth]{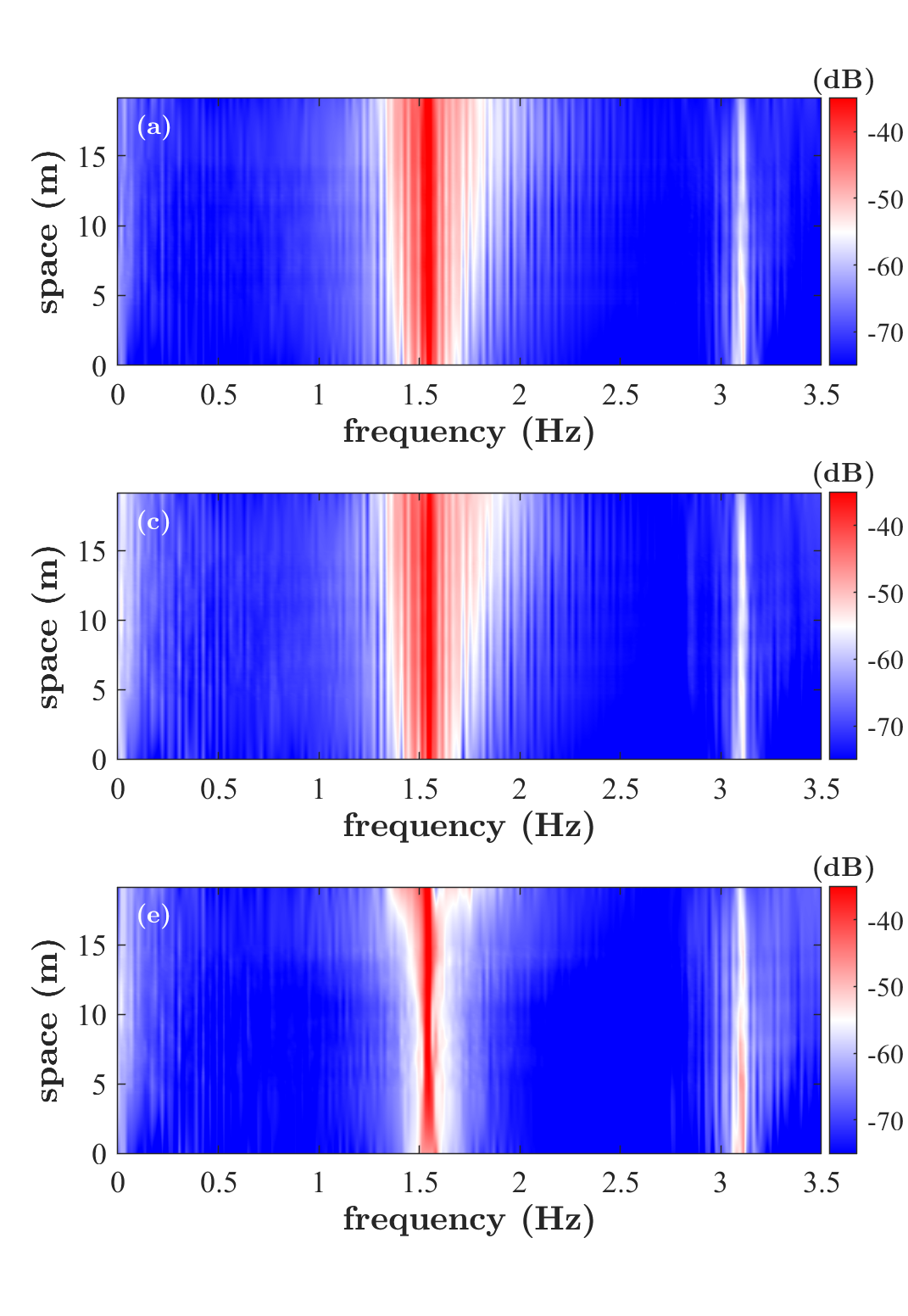}
\includegraphics[width=0.23\textwidth]{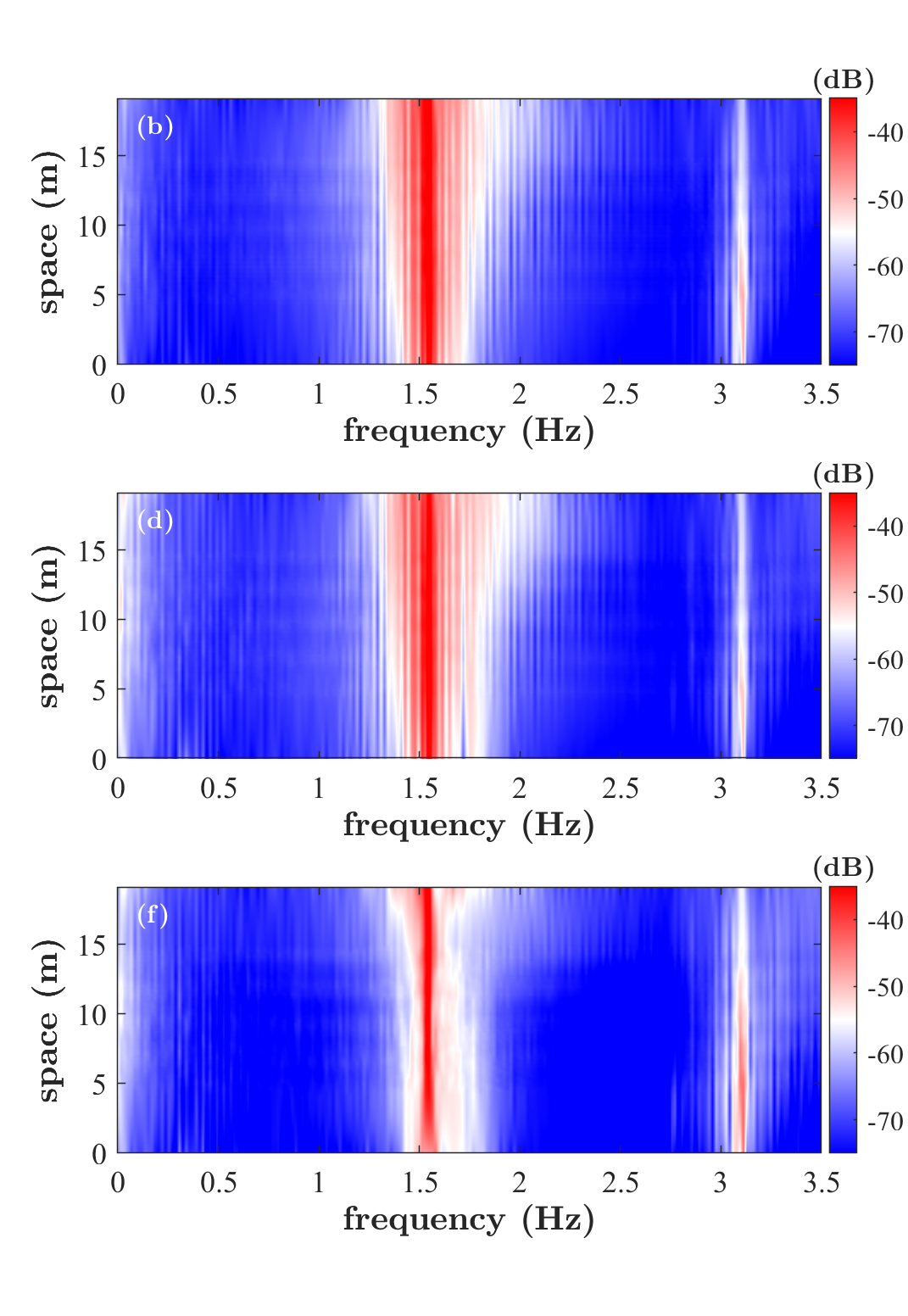}
\caption{Spectral evolution of the standing Peregrine breathers, reported in Fig. \ref{fig2}. (a) Computed for the measured surface elevation for the carrier parameters $a=0.01$ m and $ak=0.09$. (c) Incident PB wave dynamics isolated from (a). (e) Opposing wave train as isolated from (a). (b) Computed for the measured surface elevation for the carrier wave parameters $a=0.01$ m and $ak=0.10$. (d) Incident PB wave dynamics isolated from (b). (f) Opposing wave train as isolated from (b).}
\label{fig3}
\end{figure}

The wave tank measurements show an excellent agreement with the CNLSE dynamics: the PB remarkably evolves in standing wave states without any signs of disintegration while keeping its coherence. Moreover, the characteristic amplitude amplification of four in this collinear case is also reached at the expected location in the water wave tank. This wave interaction can be also confirmed when studying the respective bi-spectral evolution, see Fig. \ref{fig3}.

%\begin{figure}[H]
%\includegraphics[width=0.48\textwidth]{figure/peregrine spectrum.png}
%\caption{(a) Experimental and (b) numerical power spectrum of the standing PB evolution for $a=0.01$ m and $ak=0.09$ over 18.75 m.}
%\label{fig3}
%\end{figure}

Here, the decomposition of incident and reflected wave constituents in the bi-spectra, which are estimated from the Fourier components \cite{goda1977estimation}, prove that the extreme wave focusing only occurs for incident waves, dominated by the PB dynamics. To be more precise, the spectral broadening, which is an indicator of physical wave focusing, occurs only in the incident wave field, as in Fig. \ref{fig3} (c) and (d). On the other hand, the energy of the counter-propagating waves remains steady, see Fig. \ref{fig3} (e) and (f). We can also notice that the second harmonic energy weakens when the standing wave field has been fully developed. 
%Considering the regular amplitude of reflected wave field, the characteristic maximal wave amplitude is four times the initial amplitude. Due to the small spatial resolution, we were also able to reconstruct the spatial-series from the time-series. This is depicted in Fig. \ref{fig4}.
%\begin{figure}[H]
%\includegraphics[width=0.49\textwidth]{figure/peregrine peak results.png}
%\caption{(a) and (c) Maximal standing PB focusing as the %time-series. (b) and (d) reconstructed space-series of (a) and (c), respectively.}
%\label{fig4}
%\end{figure}

Note that the variations in surface wave profiles are very sensitive to the single gauge location. On a node point, the amplitude of the standing regular wave field is zero \cite{aguilera2021localized}. It is also worth mentioning that the PB did not have any influence in the destabilization of the regular opposing wave field. 
This persistence may be partly explained by the form of the CNLSE (\ref{cnls}), which supports conservation of integrals $\int_{-\infty}^{\infty}{|\psi_j|^2 dx}$ for $j=1,2$, so that the nonlinear exchange terms in Eqs. (\ref{cnls}) lead to frequency corrections only. The impossibility of energy exchange between two counter propagating planar deep-water wave systems due to nonlinear four-wave interactions beyond the narrow-band approximation was recently emphasized \cite{Dyachenkoetal2017,Dyachenko2020}. Nonetheless, the generation of collinear opposite waves was observed in fully nonlinear hydrodynamic simulations in \cite{Slunyaev2018}.

%\as{???We suspect that this is the effect of the negligible role of higher-order bound waves in standing wave conditions.???}

\begin{figure}[H]
\begin{subfigure}{0.48\textwidth}
\includegraphics[width=1\textwidth]{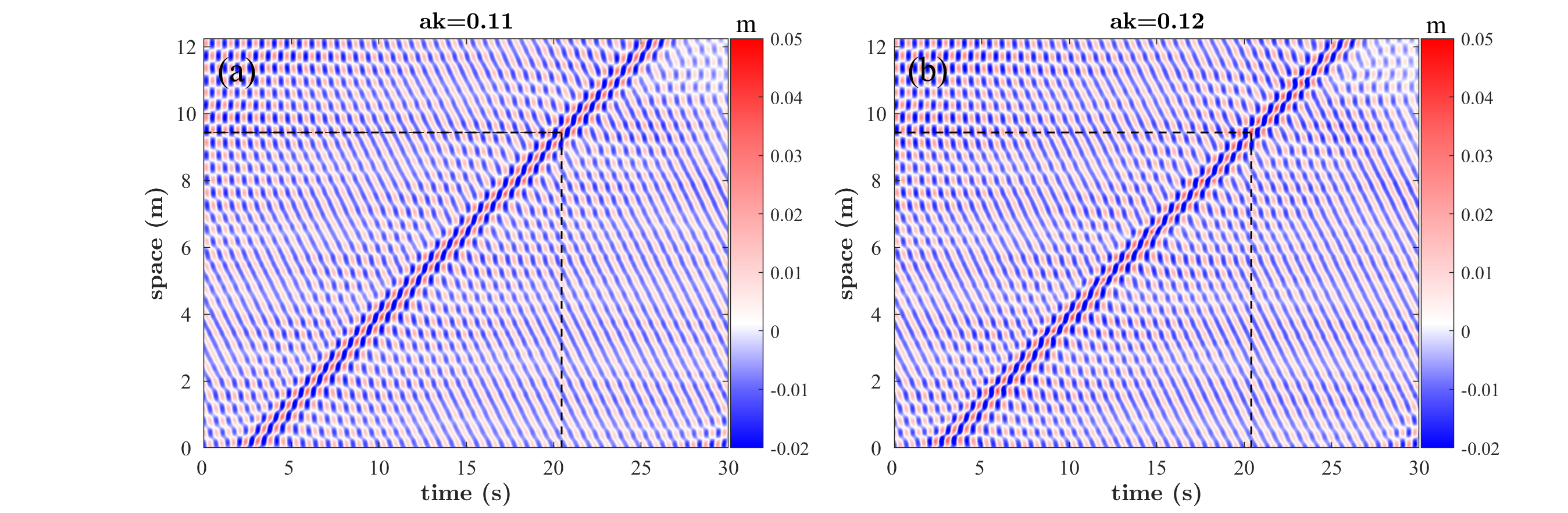}
\end{subfigure}
\begin{subfigure}{0.48\textwidth}
\includegraphics[width=1\textwidth]{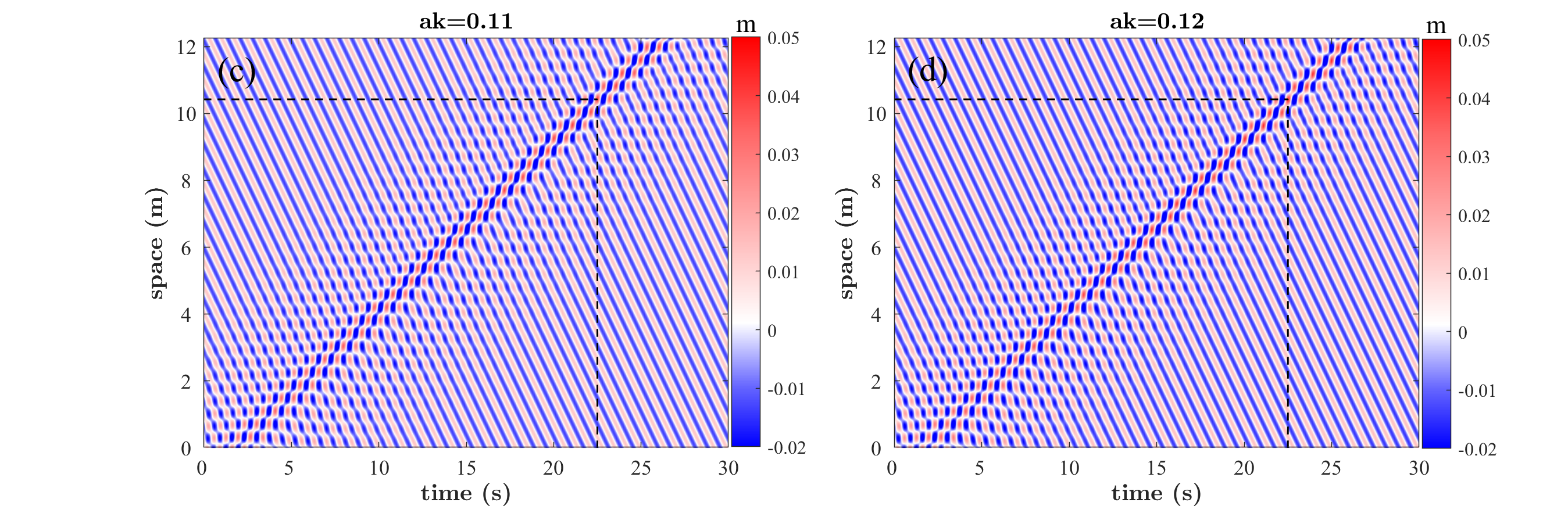}
\end{subfigure}
\caption{Top (a) and (b): Evolution of two standing degenerate solitons. Bottom (c) and (d): Corresponding CNLSE simulations. Left (a) and (c): Carrier wave parameters $a=0.01$ m  and $ak=0.11$. Right (b) and (d): Carrier wave parameters $a=0.01$ m and $ak=0.12$.}
\label{fig4}
\end{figure}

In the following we investigate the evolution of the PB counterpart on zero background \cite{kedziora2014rogue,chabchoub2021peregrine}, i.e. the degenerate soliton solution, in the presence of a counter-propagative regular wave field. Because this particular NLSE solution has finite length construction on a zero background, the standing wave patterns appear only locally during the interaction with the opposing wave field. Experimental data and respective CNLSE simulations are displayed in Fig. \ref{fig4}. 

Once again, an excellent agreement can be noticed between laboratory experiments and numerical simulations, suggesting an elastic collision between the regular envelope and degenerate soliton \cite{dyachenko2020canonical} and by that confirming once again that the CNLSE framework may have an extended applicability range beyond the physical limitations technically limiting its exploitation. 

\begin{figure}[H]
\includegraphics[width=0.48\textwidth]{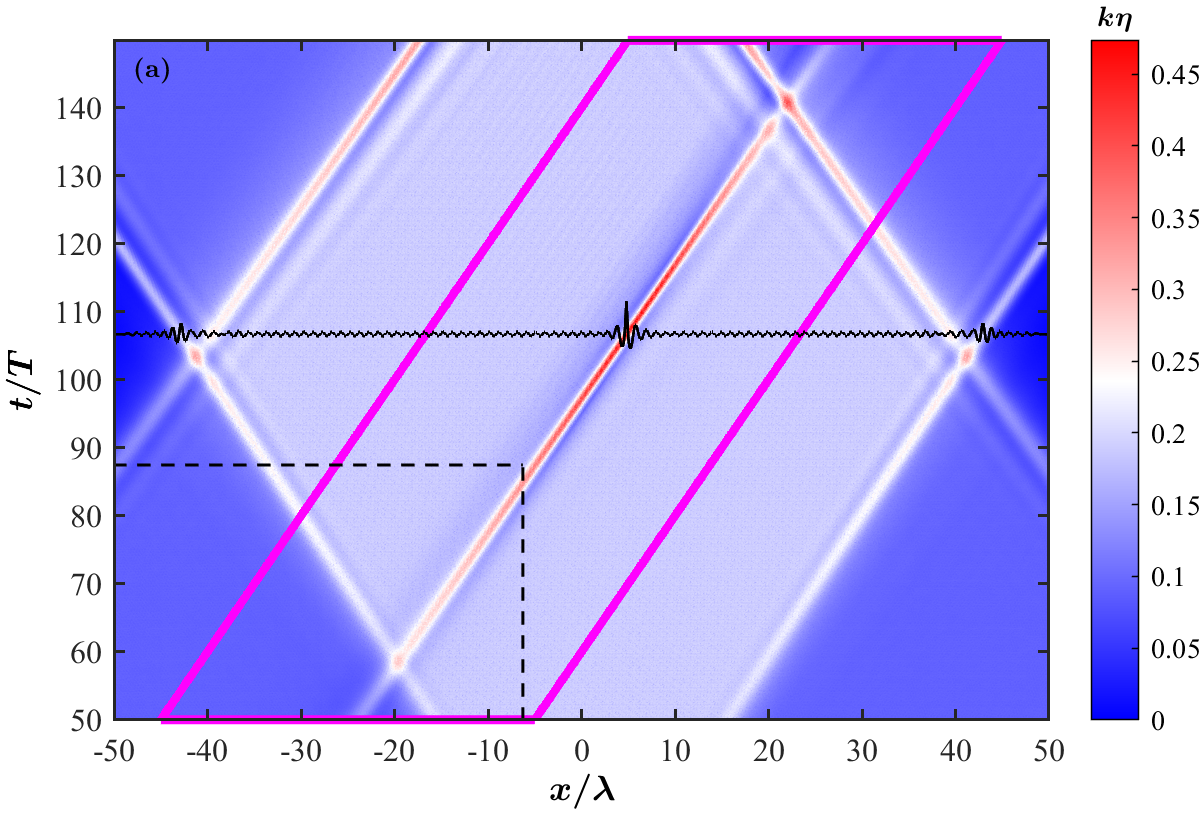}
\includegraphics[width=0.515\textwidth]{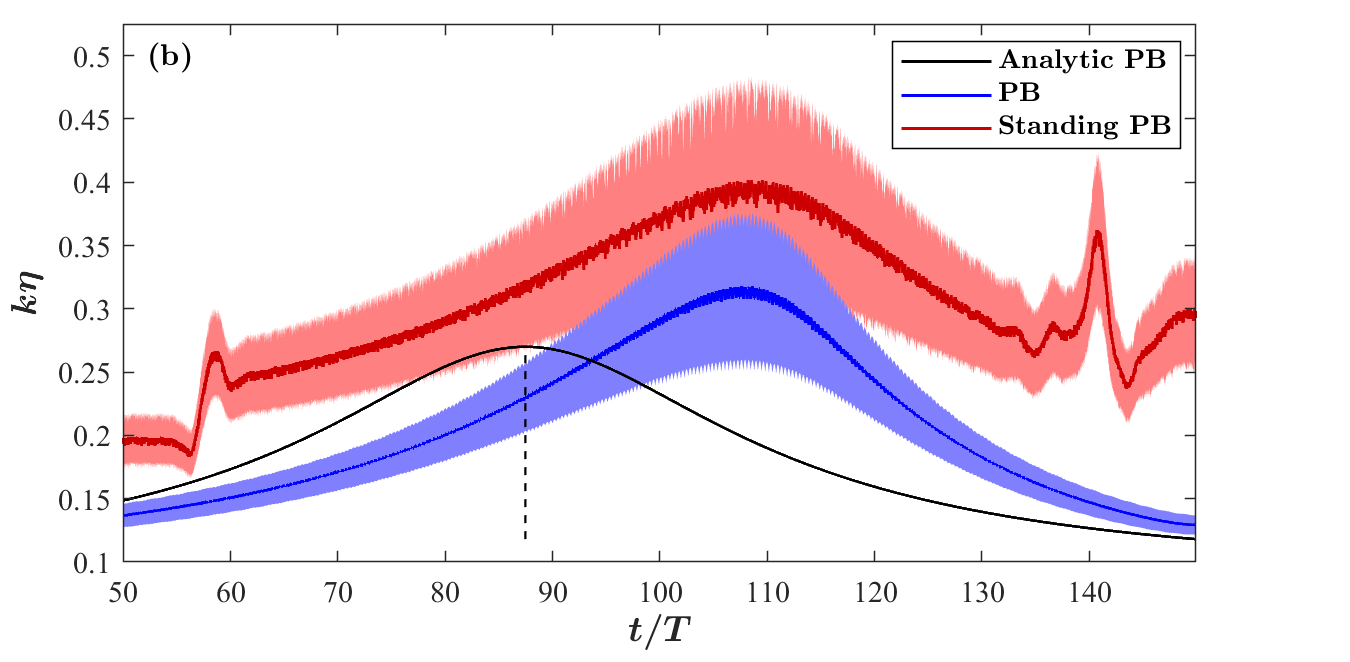}
\caption{Evolution of a PB in standing waves with the carrier steepness $ak=0.09$ as simulated by the HOSM by arranging two dozen phases allowing for the reconstruction of wave envelope. (a) The surface displacement envelope $k\eta$ is shown by the pseudo-color. The surface displacement profile of the highest focused wave is displayed in black while the dashed black lines point the location of maximal envelope compression, both according to the analytic PB solution. The magenta parallelogram shows the area where the maximum wave amplitude is evaluated. The coordinate and time are normalized with the dominant wave length $\lambda$ and period $T$, respectively. (b) Evolution of the maximum wave amplitude in the HOSM simulations and according to the analytic PB solution of the NLSE. The filled areas correspond to the intervals between crest and trough amplitudes, while the solid curves depict their simple means.}
\label{fig5}
\end{figure}

As next, we extend our proof of concept validation study in considering fully nonlinear water wave framework by numerically solving the Euler equations using the HOSM \cite{Westetal1987,dommermuth1987high}, which resolves up to $7$-wave interactions in our simulations. This approach does not only allow consideration of the wave evolution for longer time and distance than at disposal in experimental wave facilities, but also provides a higher reliability compared to weakly nonlinear CNLSE approach. Fig.~\ref{fig5} shows a corresponding and particular case study, which has been experimentally studied and reported in Fig. \ref{fig2} (a) and (c), i.e. the case of the Peregrine breather on finite background. The initial condition at $t=0$ is specified in the form of two long trains with the carrier wave steepness $ak=0.09$, and which travel towards each other. The rightward moving train is produced from the analytic PB solution; the inoculating perturbation is characterized by the steepness less than $0.11$.

The pseudo-color in Fig.~\ref{fig5} (a) represents the evolution of the surface displacement envelope, which is produced from two dozen simulations when various phase combinations were assigned to the initial wave trains. Hence, the envelope corresponds to the phase averaging of the upper enveloping surface. Only a part of the simulated domain is shown. The target focusing time according to the exact NLSE PB solution is about $90$ wave periods, see the dashed black lines. 

The snapshot of the maximum wave, which is very close to the breaking onset, is given by the black curve. Note that the deviations of the maximal envelope compression location are less significant in the wave flume because of the shorter propagating distance considered, a constraint imposed by the wave generator frequency range and the limited length of state of the art wave facilities.  

The strongly nonlinear simulations reproduce all the main features of the localized standing wave patterns as observed in the laboratory environment, particularly the clean and quasi-undisturbed extreme wave focusing within the incident PB group. At the same time, the long evolution reveals some distinctions. Differently than the weakly nonlinear framework, the strongly nonlinear simulation results correspond to slightly faster movement of the growing modulation and noticeably longer focusing time. Besides, a minor asymmetry of the emerged large wave group is clearly seen before and after the maximal wave focusing.

Evolution of the corresponding maximum wave amplitude is displayed in Fig.~\ref{fig5} (b) by the red color. It is evaluated within the area bounded by the magenta contour in Fig.~\ref{fig5} (a), which has been introduced to reduce the effect of emerging large modulations at the edges of the colliding wave trains. This dependence is compared with the result of a similar strongly nonlinear numerical simulations of the PB when the opposite wave train is absent (blue color) and with the exact analytic PB solution (black curve). The differences between the red and blue curves correspond well to the amplitude of the opposite wave train at $ak=0.09$, confirming once again the non-influence of the oppositely propagating regular wave train on the focusing dynamics of the modulationally unstable wave packet. The strongly nonlinear localized focusing of Peregrine breathers on top of progressive or standing waves occurs later than predicted by CNLSE and results in significantly larger focused waves. A behaviour which has been also quantified in the uni-directional NLSE case \cite{Slunyaevetal2013,shemer2013peregrine}.

Conclusively, we have reported experimental evidence of quasi-unperturbed PB hydrodynamics in standing waves. The same has been observed for the degenerate soliton, which is the counterpart of PB on zero background. Our results confirm that breather solutions of the NLSE can be considered to model and describe extreme wave localizations in non-integrable systems \cite{kivshar1989dynamics,chekhovskoy2019nonlinear}. Since experiments are always subject to dissipative effects, we believe that the role of NLSE solitons and breathers can be extended to a wider range of complex systems \cite{haelterman1992dissipative,akhmediev2008dissipative,nielsen2021nonlinear}. We also anticipate that our experimental work will motivate further studies to investigate the role of breathers for standing wave conditions in nonlinear dispersive media as well as to determine their function in crossing seas and vector nonlinear fiber-analog systems.  

A.S.  acknowledges the support from the RFBR grant No. 21-55-15008 and from  Lab. of Dynamical Systems and Applications NRU HSE (the Ministry of  Science  and  Higher Education of the Russian Federation Grant No. 075-15-2019-1931). A.C. acknowledges support from Kyoto University's Hakubi Center for Advanced Research. 

\bibliography{reference}

\end{document}